
\documentstyle{amsppt}
\input amssym.def
\loadeusm

\magnification=\magstep1
\NoBlackBoxes

\topmatter
hep-th/9308109 , IMA Preprint Series no.1153, July 1993
\title{QUADRICS ON COMPLEX RIEMANNIAN SPACES\\
OF CONSTANT   CURVATURE, SEPARATION \\
 OF VARIABLES AND THE GAUDIN  MAGNET}
\endtitle

\author
E.G. KALNINS,  V.B. KUZNETSOV\footnotemark"*"
and  WILLARD MILLER, Jr.$^{\dag}$
\endauthor

\affil
Department of Mathematics and Statistics\\
 University of Waikato\\
 Hamilton, New Zealand\\
\\
Department of Mathematics and Computer Science\\
 University of Amsterdam\\
 Plantage Muidergracht 24  \\
 1018 TV Amsterdam          \\
 The Netherlands.             \\
\\
School of Mathematics\\
 and Institute for Mathematics and its Applications,\\
 University of Minnesota,\\
Minneapolis, Minnesota  55455, USA. \\
\endaffil

\thanks{*This author was supported by the
National Dutch Science
Organization (NWO) under the Project \# 611--306--540.
\hfill\newline
$\dag$ Work  supported in part by the National Science
Foundation under grant DMS 91--100324}
\endthanks

\leftheadtext{E.G. KALNINS, V.B. KUZNETSOV AND WILLARD MILLER, Jr.}
\rightheadtext{SEPARATION  OF VARIABLES AND THE GAUDIN  MAGNET}

\abstract{
We consider integrable systems that are connected with
orthogonal separation of variables in complex Riemannian spaces of constant
curvature. An isomorphism  with the hyperbolic
Gaudin magnet, previously pointed out by one of us,
 extends to coordinates of this type. The complete classification
of these separable coordinate systems is provided by means of the
corresponding $L$-matrices for the Gaudin magnet. The limiting procedures (or
$\epsilon $ calculus)
which relate various degenerate orthogonal coordinate systems play a crucial
result in the classification of all such systems.}
\endabstract
\endtopmatter

\document

\subheading{1. Classical Integrable Systems on Complex Constant Curvature
Spaces
and the Complex Gaudin Magnet}
Separation of variables in the Hamilton Jacobi equation
\medskip
$$
H(p_1,...,p_n;x_1,...,x_n)= \sum ^n_{\alpha ,\beta  =1}
g^{\alpha \beta }p_\alpha p_\beta =E,\quad p_\alpha =
{\partial W\over \partial x_\alpha },\quad\alpha  =1 ,...,n \tag1.1
$$
amounts to looking for a solution of the form
$$
W = \sum ^n_{\alpha  =1} W_\alpha (x_\alpha ;h_1,...,h_n),\quad h_n=E\tag1.2
$$
The solution is said to be a {\it complete integral}\ if
$\det (\partial ^2W/\partial x_i\partial h_j)_{n\times n}\neq 0$. The solution
then describes that of free motion on the corresponding Riemannian space with
contravariant metric $g^{\alpha \beta }$. Indeed, if we require $b_j= {\partial
W\over \partial h_j }-h_nt
\delta_{nj},
\quad j =1 ,...,n$ for parameters $b_j$, we find that the functions
${\bold x}({\bold b}, {\bold h})$,
${\bold p}({\bold b}, {\bold h})$ satisfy Hamilton's equations
$$
{\dot x}_\alpha=\frac{\partial H}{\partial p_\alpha},\quad {\dot
p}_\alpha=-\frac{\partial H}{
\partial x_\alpha}\tag1.3
$$
 In this article we allow the
Riemannian space to be complex and we consider variable separation of (1.1) on
the
following two classes of spaces.

\roster
\item The $n$ dimensional complex sphere $S_{n{\bold C}}$. This is commonly
realised
by the set of complex vectors ${\bold x} = (x_1,...,x_{n+1})$ which satisfy
$\sum ^{n+1}_{\alpha =1} x^2_\alpha  =1$ and have infinitesmal distance
$d{\bold x}\cdot d{\bold x}
= \sum ^{n+1}_{\alpha =1} dx^2_\alpha $

\item The $n$ dimensional complex Euclidean space $E_{n{\bold C}}$. This is the
set of
complex vectors ${\bold x} =(x_1,....,x_n)$ with infinitesmal distance $d{\bold
x}\cdot d{\bold x}=
\sum ^n_{\alpha =1}dx^2_\alpha $.
\endroster
A fundamental problem from the point of view of separation of variables on
these manifolds is to find all ``inequivalent'' coordinate systems. As yet,
this is an
 unsolved problem, principally because many such coordinate systems are
intrinsically nonorthogonal. For orthogonal coordinate systems the problem is
completely solved and in this case the constants $h_i$ occuring in the complete
integral can be
chosen to be the values of an involutive set of constants of motion
$$A_j= \sum ^n_{\alpha ,\beta  =1} a^{(j)}_{\alpha \beta }p_\alpha p_{\beta},
\quad j=1,...,n,\  A_n=H,\tag1.4
$$
$$\{A_j,A_k\}=0,
$$
where
$$
\{F(x_\alpha  ,p_\beta ),G(x_\alpha  ,p_\beta )\} = \sum ^n_{\gamma =1}
({\partial F\over \partial p_\gamma } {\partial G\over \partial x_\gamma }-
{\partial F\over \partial x_\gamma }{\partial G\over \partial p_\gamma }),
\quad \alpha ,\beta =1,...,n
$$ is the Poisson bracket.  These constants of the
motion are such that

\roster
\item each of the tensors $a^{(j)}_{\alpha \beta }$ is a Killing tensor and
satisfies Killing's equations \hfill\break
$\nabla _{(\alpha }a^{(j)}_{\beta \gamma )} =0$,
\cite{1}, and

\item
the $A_j$ can be represented as a sum of quadratic elements of the
enveloping algebra of the Lie algebra of symmetries of each of these two
considered spaces.
\endroster
The Lie algebras of these spaces have respectively bases of the form

\roster
\item $SO(n+1): M_{\alpha \beta }=x_\alpha p_\beta -x_\beta p_\alpha,
\quad \alpha ,\beta  =1,...,n+1$. Here,
 $$
 \{M_{\alpha \beta }, M_{\gamma \delta }\}
=\delta _{\alpha \gamma }M_{\delta \beta }+\delta _{\alpha \delta }M_{%
\beta \gamma }+\delta _{\beta \gamma }M_{\alpha \delta }+\delta _{\beta \delta
}M_{\gamma \alpha }
$$

\item
$E(n): M_{\alpha \beta }, P_\gamma=p_\gamma,\  \alpha,\beta,\gamma=
1,\cdots,n,\ \alpha\ne
\beta$.
Here,
$$\{M_{\alpha \beta }, P_\gamma \}
=\delta _{\beta \gamma }P_\alpha -\delta _{\alpha \gamma }P_\beta,\ \{P_\alpha
,P_\beta \} =0.$$
\endroster

In this section the separable coordinate systems classified in \cite{1,5,10}
 are given
 an algebraic interpretation. This is done using the complex analogue of
the isomorphism between all integrable systems connected with all possible
separable systems and the $m$ site $SO(2,1)$ Gaudin magnet, \cite{2,4}. The
$m$-site
complex Gaudin magnet can be realised as follows. Consider the direct sum of
Lie algebras, each of rank 1,
$$ {\Cal A} = \oplus ^m_{\alpha =1} so_\alpha (3,{\bold C}).\tag1.5$$
The generators ${\bold s}_\alpha  \in \ {\bold C}^3,\ \alpha  =1,\cdots,m$ of $
\Cal A$
satisfy the Poisson bracket relations

$$\{s^i_\alpha  ,s^j_\beta \}
=-\delta _{\alpha \beta }\epsilon _{ijk}s^k_\alpha.\tag1.6$$

The following metric will be used subsequently when norms and scalar
products are calculated:
$${\bold s}^2_\alpha  = ({\bold s}_\alpha  ,{\bold s}_\alpha  ) =
(s^1_\alpha )^2+(s^2_\alpha )^2+(s^3_\alpha )^2,\quad
({\bold s}_\alpha  ,{\bold s}_\beta  ) =
s^1_\alpha s^1_\beta +s^2_\alpha s^2_\beta +s^3_\alpha s^3_\beta. $$
If for each $\alpha$, ${\bold s}^2_\alpha  = c^2_\alpha$ then the variables
${\bold s}_\alpha $
lie on the direct product of $n$ complex spheres in ${\bold C}^3$. The
complex  Gaudin magnet is the integrable Hamiltonian
system described by the $n$ integrals of motion $H_\alpha$ which are in
involution with respect to the Poisson bracket,
$$H_\alpha =2\ {\sum ^{m}_{\beta =1}}'{({\bold s}_\alpha  ,{\bold s}_\beta
)\over e_\alpha
-e_\beta } ,\quad \{H_\alpha  ,H_\gamma \}=0.\tag1.7$$
(We will give a simple proof of this involution property later.)
Here the $e_\alpha $ are taken to be pairwise distinct. This integrable
Hamiltonian system
is called an {\it $m$-site $SO(3,{\bold C})-$XXX Gaudin magnet}. The $H_\alpha$
are all
quadratic functions in the generators of the $\Cal A$ algebra and the following
identities are satisfied:
$$\sum ^m_{\alpha =1}H_\alpha =0,\quad \sum ^m_{\alpha =1}e_\alpha H_\alpha  =
{\bold J}^2 -
\sum ^m_{\alpha =1}c^2_\alpha, \tag1.8 $$
where we have introduced the variables
$${\bold J} = \sum ^m_{\alpha =1}{\bold s}_\alpha,\quad {\bold J}^2 =({\bold
J},{\bold J})
,\tag1.9 $$
the total sum of the  momenta ${\bold s}_\alpha$.
Indeed
$$\{J^i,J^j\} = -\epsilon _{ijk}J^k,\quad \{J^k,H_\alpha \} =0,\quad
i,j,k=1,2,3,\ \alpha
=1,\cdots,m.\tag1.10$$

The complete set of involutive integrals of motion is provided by $H_\alpha
,{\bold J}^2$ and, for example, $J^3$. The integrals are generated by the
$2\times 2\ L$
matrix \cite{3,4,6,7,8},
$$L(u) = \sum ^m_{\alpha =1}{1\over u - e_\alpha }\left( \matrix s_\alpha^3 &
-s^1_\alpha - is^2_\alpha\\ -s^1_\alpha + is^2_\alpha & -s_\alpha^3 \endmatrix
\right)= \left(\matrix A & B\\C & -A\endmatrix \right),\tag1.11
 $$
where
$$\det L(u) = -(A^2+BC)=-\sum ^m_{\alpha =1}{H_\alpha \over u - e_\alpha } -
\sum ^m_{\alpha =1}{c^2_\alpha \over (u - e_\alpha )^2}.\tag1.12 $$
Furthermore, $L(u)$  satisfies the linear $r$ matrix algebra  \cite{2,21}
$$
\{ {\overset 1 \to L}(u), {\overset 2 \to L}(v)\}=\frac{i}{u-v}
[P, {\overset 1 \to L}(u)+ {\overset 2 \to L}(v)],\quad i=\sqrt{-1}\tag1.13
$$
where
$$P=\left(\matrix 1 & 0 & 0& 0\\
0& 0& 1& 0\\
0& 1& 0& 0\\
0& 0& 0& 1\endmatrix \right),\quad
{\overset 1 \to L}(u)=L(u)\otimes I,\ {\overset 2 \to L}(v)=I\otimes
L(v).\tag1.14
$$
The algebra specified by (1.13)   contains all the necessary
Hamiltonian structure of the problem in question. Note that (1.13) is
equivalent
to the easily proved relations
$$ \align
\{A(u),A(v)\}&= \{B(u),B(v)\}= \{C(u),C(v)\}=0,\tag1.15 \\
\{A(u),B(v)\}&=\frac{i}{u-v}(B(v)-B(u)),\quad
\{A(u),C(v)\}=\frac{i}{u-v}(C(u)-C(v)),\\
\{B(u),C(v)\}&=\frac{2i}{u-v}(A(v)-A(u)).
\endalign$$
{}From (1.15) and the Leibnitz property of the Poisson bracket
it is straightforward to deduce that
$$
\{\det L(u),\det L(v)\} =0.$$
 In his article \cite{4} Kuznetsov
has explicitly given the nature of the isomorphism between the XXX Gaudin
magnet models and the separation of variables on the $n$ dimensional real
sphere
$S_n$.  The purpose of this article is to extend
these ideas to complex orthogonal coordinate systems on the complex $n$ sphere
$S_{n{\bold C}}$ and, of course, as a consequence complex Euclidean space
$E_{n{\bold C}}$. Following Kuznetsov \cite{4,7} in the case of the sphere, we
set
$c_\alpha =0,\ \alpha =1,\cdots,n+1$. The coordinates on the resulting cones
are
parametrised by
$$ s^1_\alpha = {1\over 4} (p^2_\alpha +x^2_\alpha ),\quad s^2_\alpha = {i\over
4}
(p^2_\alpha -x^2_\alpha ),\quad s^3_\alpha = {i\over 2} p_\alpha
x_\alpha.\tag1.16
$$
It follows from (1.6) that $\{x_\alpha ,x_\beta \}=\{p_\alpha ,p_\beta \}=0$,
 $\{p_\alpha ,x_\beta \}=\delta _{\alpha \beta }$. Introducing the new
variables $M_{\alpha \beta }=x_\alpha p_\beta - x_\beta p_\alpha $ which are
the generators of rotations we have
$$ ({\bold s}_\alpha, {\bold s}_\beta )= {1\over 8} M^2_{\alpha \beta }.$$
These generators satisfy the commutations relations given previously. This
equality establishes the simple quadratic connection between the generators
${\bold s}_\alpha $ of ${\Cal A}$ and the $M_{\alpha \beta }$ of $SO(n+1)$.
Under this
isomorphism the integrals given in (1.7) (and the subsequent discussion)
   transform into the following
integrals for the free motion on the $n$ sphere
$$H = \sum ^{n+1}_{\alpha =1}h_\alpha H_\alpha  = {1\over 4} \sum _{\alpha
<\beta }
{h_\alpha -h_\beta \over e_\alpha -e_\beta } M^2_{\alpha \beta }.$$
For $h_\gamma =e_\gamma $ we obtain the Casimir element of the $so(n+1)$
algebra $\sum _{\alpha <\beta }M^2_{\alpha \beta }$. The total
momentum $\bold J$ takes the form
$$J^1= {1\over 4}({\bold p}\cdot {\bold p} + {\bold x}\cdot {\bold  x}),
\quad J^2= {i\over 4}({\bold p}\cdot{\bold  p} - {\bold x}\cdot {\bold x}),
\quad J^3=
{i\over 2} {\bold p}\cdot {\bold x}$$
where the scalar product for the vectors $\bold x$ and $\bold p$ in ${\bold
C}^{n+1}$ is
Euclidean.The quantities $\bold M$  and $\bold J$ form the direct sum
$so(n+1)\oplus so(3)$
as a result of the commutation relations
$$\{M_{\alpha \beta },J^i\} =0 ,\quad \{J^i,J^j\} = -\epsilon _{ijk}J^k.$$
Therefore, in addition to the involutive set of integrals $H_\alpha $ we can
choose ${\bold J}^2 = {1\over 4} \sum _{\alpha <\beta }M^2_{\alpha \beta }$ and
$2(J^1+iJ^2)={\bold x}\cdot{\bold x}=c$, which gives the equation of the $n$
sphere.

\subheading{2. Generic Ellipsoidal Coordinates on $S_{n{\bold C}}$ and
 $E_{n{\bold C}}$}
Critical to the separation of variables on the $n$ sphere $S_{n{\bold C}}$ is
the
system of ellipsoidal coordinates graphically pictured by the
irreducible block
$$(S_{n{\bold C}}|e_1|e_2| \cdots |e_{n+1}|.\tag2.1$$
where in general $e_\alpha  \neq  e_\beta$ for $\alpha\ne \beta$. The
separation variables are
defined as zeros of the off diagonal element $B(u)$ of the $L$
matrix, i.e., $B(u_j)=0,\ j=1,...,n$. It follows that
$$\sum ^{n+1}_{\alpha =1}\frac{x^2_\alpha}{u - e_\alpha} = 0\ \  \text{for}\ \
u=u_j
\ \text{and}\  \
x^2_\alpha  = c
{\Pi ^{n}_{j=1}(u_j-e_\alpha )\over \Pi _{\beta \neq \alpha }(e_\beta -e_%
\alpha )}.\tag2.2
$$
Each vector of momentum ${\bold s}_\alpha $ is associated with a cell $e_\alpha
$ of
the block. Note that $x^2_\alpha  = 2(s^1_\alpha +is^2_\beta)$. For each
$u_j$ the conjugate variable $v_j$ is defined according to
$$v_j=-iA(u_j)= {1\over 2}
\sum ^{n+1}_{\alpha =1}{x_\alpha p_\alpha \over u_j-e_\alpha }.\tag2.3$$
{}From (1.15) one can show that the $u_i, v_i$ satisfy the  canonical relations
\cite{4}:
$$
\{u_j,u_i\}=\{v_j,v_i\}=0,\  \ \{v_j,u_i\}=\delta _{ij}.\tag2.4
$$
The change to
the new variables $v_j,u_i,c$ and $J^3$ is effectively the procedure of
variable separation of (1.1) in ellipsoidal coordinates
on the $n$-sphere. Writing the $L$ matrix in terms of the new variables we
obtain
$$
L(u)=\left( \matrix A(u)& B(u)\\ C(u)& -A(u)\endmatrix \right),
\quad B(u)=- {c\over 2}
{\Pi ^{n}_{j=1}(u-u_j)\over \Pi ^{n+1}_{\alpha =1}(u-e_\alpha )},
$$
$$
A(u)=-\frac{2i}{c}B(u)\left(-iJ^3+ \sum ^{n}_{j=1}{v_j\over u-u_j}
{\Pi ^{n+1}_{\gamma =1}(u_j-e_\gamma )\over \Pi _{i\neq
j}(u_j-u_i)}\right),\tag2.5
$$
where
$A(u_j)=iv_j$, $j=1,\cdots, n$, and $A(u)\rightarrow (1/u)J^3+\cdots$ as
$u\rightarrow \infty$.

To obtain $C(u)$ we first notice that equating residues at $e_\alpha $ on the
right and left hand side of $A(u)$ gives
$$
p_\alpha =\frac{2x_\alpha }{c}\left(-iJ^3+ \sum ^{n}_{j=1}{v_j\over
e_\alpha-u_j}
{\Pi ^{n+1}_{\gamma =1}(u_j-e_\gamma )\over \Pi _{i\neq j}(u_j-u_i)}\right).
$$
This together with the expression (2.2) for $x^2_\alpha $ in terms
of $u_j$ gives $C(u)$ in the new
variables. Three other useful formulae are
$$
M_{\alpha \beta } = {2x_\alpha x_\beta \over c} (e_\alpha -e_\beta )
\sum ^{n}_{j=1} v_j
{\hat \Pi^{n+1}_{\gamma =1}(u_j-e_\gamma )\over \Pi _{i\neq
j}(u_j-u_i)},\tag2.6
$$
where the hat in (2.6) means that the product terms with $\gamma=\alpha$
and $\gamma=\beta$ are omitted,
$$
{\bold J}^2 =- \sum ^{n}_{j=1} v^2_j
{ \Pi ^{n+1}_{\gamma =1}(u_j-e_\gamma )\over \Pi _{i\neq j}(u_j-u_i)} =
\sum ^{n}_{j=1} cv^2_j\left(
\sum ^{n+1}_{\alpha=1}{x^2_\alpha \over (u_j-e_\alpha )^2}\right)^{-1},\tag2.7
$$
$$H_\alpha = \frac{x^2_\alpha}{c}\sum ^{n}_{j=1}{v^2_j\over u_j-e_\alpha}
{\Pi ^{n+1}_{\gamma =1}(u_j-e_\gamma )\over \Pi _{i\neq j}(u_j-u_i)}.
$$
\medskip
\noindent
These relations together with (2.2)  establish the explicit connection between
the
two sets of $2n+2$ variables $p_\alpha ,x_\alpha $ and $u_i,v_j,c$, $J^3$. The
equation for the eigenvalue curve $\Gamma :\det (L(u)-i\lambda I)=0$ has the
form
$$-\lambda ^2-A(u)^2-B(u)C(u)=0.$$
If we put $u=u_j$ into this equation then $\lambda =\pm v_j$. Thus variables
$u_j$
and $v_j$ lie on the curve $\Gamma $:
$$v^2_j+ \sum ^n_{\alpha =1} {H_\alpha \over u_j-e_\alpha }\equiv v^2_j-\det
L(u_j) =0.\tag2.8
$$

Equations (2.8) are the separation equations for each of the $n$
degrees of freedom connected with the values of the integrals $H_\alpha$. For
the sphere $c_\alpha =0$ these have the form
$$H_\alpha = {1\over 4} {\sum_\beta}'
{M^2_{\alpha \beta }\over e_\alpha -e_\beta },\quad \sum _\alpha H_\alpha =0,
\quad
\sum _\alpha e_\alpha H_\alpha = {1\over 4}
\sum _{\alpha <\beta }M^2_{\alpha \beta } ={\bold J}^2.
$$
The Hamilton Jacobi equation (1.1) when parametrised by these variables has the
form
$$
{1\over 4} \sum _{\alpha <\beta }M^2_{\alpha \beta }
={\bold J}^2=-\sum ^{n}_{j=1}(\frac{\partial W}{\partial u_j})^2
{\Pi ^{n+1}_{\gamma =1}(u_j-e_\gamma )\over \Pi _{i\neq j}(u_j-u_i)} =E,\quad
v_j=\frac{\partial W}{\partial u_j},\tag2.9
$$
which can be solved  by the separation of variables ansatz
$$W = \sum ^{n}_{j=1}W_j(u_j,H_1,...,H_n)= \sum ^{n}_{j=1}\int \
v_jdu_j.\tag2.10
$$

It will be convenient to employ an alternative form of the $L$ matrix. If we
use
the vector
$${\bold L}(u)= \sum ^{n+1}_{\alpha =1}{{\bold s}_\alpha \over u-e_\alpha },
\quad L(u)=({\bold s}, {\bold L}(u)),\quad \det L(u)=-{\bold L}(u)\cdot{\bold
L}(u) \tag2.11
$$
where
$$s^1=\left(\matrix 0 &-1\\ -1&  0\endmatrix\right),\quad s^2
=\left(\matrix 0 &-i\\ i&  0\endmatrix\right),\quad
s^3=\left(\matrix 1 &0\\ 0&  -1\endmatrix\right),
$$
we see that $\bold L$  satisfies
$$\{L^i (u),L^j (v)\} = {\epsilon _{ijk}\over u -v} \left(L^k(u)-L^k(v)\right).
\tag2.12$$

At this point we must consider a crucial difference between the real sphere and
its complex counterpart. In the case of the complex sphere the generic
ellipsoidal coordinates can admit multiply degenerate forms: the restriction
$e_\alpha \neq  e_\beta$, for $\alpha \neq \beta $ can be lifted. The resulting
coordinates can be denoted by the block form
$$(S_{n{\bold C}}|e^{\lambda _1}_1|e^{\lambda _2}_2| \cdots |e^{\lambda _q}_q|
,\quad \lambda _1+\cdots+\lambda _q=n+1,
$$
where the $\lambda _\alpha$ denote the multiplicities of the $e_\alpha$.
To understand how
the previous analysis applies to these types of coordinates we first illustrate
with an example corresponding to the coordinates with diagram
$$(S_{n{\bold C}}|e^2_1|e_3|\cdots|e_{n+1}|.
$$
In this case we write
$$
{\bold L}(u)= {a_1{\bold s}_1\over u-e_1} + {a_2{\bold s}_2\over u-e_2} +
\sum ^{n+1}_{\alpha =3}{{\bold s}_\alpha \over u-e_\alpha }.\tag2.13
$$
Putting
$$x
_2\rightarrow  x'_1 + \epsilon x'_2,\ x_1\rightarrow  x'_1,\
p_2\rightarrow  p'_1 + \epsilon p'_2,\ p_1\rightarrow  p'_1,\tag2.14$$
$$a_1+a_2 =0,\ a_2=1/\epsilon,\  x_\alpha\rightarrow x_\alpha',\
p_\alpha\rightarrow p_\alpha',\ (\alpha\ge3),\ e_2=e_1+\epsilon,$$
 then  in the limit as $\epsilon \rightarrow
0$ we find
$$
{\bold L}(u)=\frac{{\bold z}_1}{(u-e_1)^2}+ \frac{{\bold z}_2}{ u-e_1} +
\sum ^{n+1}_{\alpha =3}\frac{{\bold z}_\alpha}{ u-e_\alpha }
$$
where
$$\align
{\bold z}_1&=<{1\over 4}({p_1'}^2 + {x'_1}^2),
{i\over 4}({p_1'}^2 -{x_1'}^2),{i\over 2}p'_1x'_1> \cr
{\bold z}_2&=<{1\over 2}(p'_1p'_2+x'_1x'_2),{i\over
2}(p'_1p'_2-x'_1x'_2),{i\over 2}(p'
_1x'_2+p'_2x'_1)>\tag2.15\cr
{\bold z}_\alpha &={\bold s}_\alpha  ,\quad \alpha  = 3,\cdots,n+1.
\endalign$$

The components of ${\bold z}_1 ,{\bold z}_2$  satisfy the $E(3,{\bold C})$
algebra  relations
$$
\{z^i_1,z^j_1\}=0,\quad \{z^i_1,z^j_2\}=-\epsilon _{ijk}z^k_1,\quad
\{z^i_2,z^j_2\}=-\epsilon _{ijk}z^k_2
$$
where the Poisson bracket (expressed in the primed coordinates) is
$$\{F,G\} =
\sum ^{n+1}_{j,k=1}B_{jk}({\partial F\over \partial p'_j}
{\partial G\over \partial x'_k}-
{\partial F\over \partial x'_j}{\partial G\over \partial p'_k})$$
 and
$$
B = (B_{jk}) = \left(\matrix 0& 1& 0& \cdots& 0 \\
 1& 0&0& \cdots&0 \\
 0& 0& 1& \cdots&0\\
  & \cdots\\
 0& 0&0&\cdots&1\endmatrix\right).
$$
It can easily be verified from these relations that relations (2.12) are again
satisfied.
Thus, $\{{\bold L}\cdot{\bold L}(u),{\bold L}\cdot{\bold L}(v)\}=0$ so the
coefficients of
the various powers $(u-e_j)^{-k}$ in the expansion of ${\bold L}\cdot{\bold
L}(u)$ form
an involutive set of integrals of motion.

We obtain
$$ \align
 {\bold L}\cdot{\bold L} &= {1\over (u-e_1)^4} {\bold z}_1\cdot{\bold z}_1+
{2\over (u-e_1)^3} {\bold z}_1\cdot{\bold z}_2 +
{1\over (u-e_1)^2} {\bold z}_2\cdot{\bold z}_2+\sum_{\alpha\ge 3}\frac{2{\bold
z}_1\cdot
{\bold z}_\alpha}{(u-e_1)^2(u-e_\alpha)} \cr
&+\sum_{\alpha\ge 3}\frac{2{\bold z}_2\cdot
{\bold z}_\alpha}{(u-e_1)(u-e_\alpha)} +
\sum _{\alpha,\beta\ge 3,\ \alpha \neq \beta } {1\over (u-e_\alpha )(u-e_\beta
)} ({\bold z}_\alpha
,{\bold z}_\beta )\cr
&= {-1\over 4(u-e_1)^2} (p'_1x'_2-p'_2x'_1)^2 +
\sum ^{n+1}_{\alpha =3}[{1\over 4(u-e_1)^2(u-e_\alpha )}
(p'_1x'_\alpha -p'_\alpha x'_1)^2 \tag2.16\cr
&+
{1\over 2(u-e_1)(u-e_\alpha )} (p'_1x'_\alpha -p'_\alpha x'_1)
(p'_2x'_\alpha -p'_\alpha x'_2)] \cr
&+
\sum ^n_{\alpha ,\beta =3,\alpha \neq \beta }
{1\over 8(u-e_\alpha )(u-e_\beta )} (p'_\beta x'_\alpha -p'_\alpha x'_\beta
)^2.
\endalign$$

In particular,
$${\bold z}_1\cdot{\bold z}_1=0,\ {\bold z}_1\cdot{\bold z}_2 =0,\ {\bold
z}_\alpha\cdot
{\bold z}_1 =
(p'_1x'_\alpha -p'_\alpha x'_1)^2/8,\ {\bold z}_2\cdot
{\bold z}_2 = -(p'_1x'_2 -p'_2 x'_1)^2/4,$$
$${\bold z}_\alpha \cdot{\bold z}_2
= (p'_1x'_\alpha-p'_\alpha x'_1)
(p'_2x'_\alpha-p'_\alpha x'_2)/4.
$$

To relate this to the projective coordinates on the complex $n$ sphere we
recall that under the transformation (2.14)
the fundamental quadratic forms $X = a_1x_1^2+a_2x_2^2+\sum
^{n+1}_{\alpha=3}x^2_\alpha$, $P =
a_1p_1^2+a_2p_2^2+\sum ^{n+1}_{\alpha=1}p^2_\alpha$ transform to
$X = 2x'_1x'_2 + \sum ^{n+1}_{\alpha=3}{x'}_\alpha^2,\
P =  2p'_1p'_2 + \sum ^{n+1}_{\alpha=3}{p'}_\alpha^2$. Therefore if we take the
coordinates
$$
x'_1={1\over \surd 2}(X_1+iX_2),\ x'_2={1\over \surd 2}(X_1-iX_2),\
x'_\alpha=X_\alpha
,\ \alpha=3,\cdots,n+1$$
$$
p'_1={1\over \surd 2}(P_1+iP_2),\ p'_2={1\over \surd 2}(P_1-iP_2),\
p'_\alpha=P_\alpha
,\ \alpha=3,\cdots,n+1
$$
we then can write
$$
p'_1x'_\alpha -p'_\alpha x'_1= {1\over \surd 2}(M_{1\alpha }+iM_{2\alpha }),\ \
p'_2x'_\alpha -p'_\alpha x'_2={1\over \surd 2}(M_{1\alpha }-iM_{2\alpha }),
$$
$$
p'_1x'_2-p'_2x'_1=i\surd 2M_{12},\ \  X =\sum ^{n+1}_{j=1}X^2_j,\ P =\sum
^{n+1}_{j=1}P^2_j,
$$
where $M_{jk}=X_jP_k-X_kP_j$.

The integrals of motion $H_\alpha  ,Z$ in this case have, using partial
fractions,
the form
$$\align
{\bold L}\cdot{\bold L}& = {1\over
(u-e_1)^2}[-\frac{1}{4}(p'_1x'_2-p'_2x'_1)^2+
\sum ^{n+1}_{\alpha =3}{1\over 4(e_1-e_\alpha) }(p'_1x'_\alpha -p'_\alpha
x'_1)^2] \cr
&+\sum ^{n+1}_{\alpha =3}{1\over u-e_\alpha }[{1\over 4(e_1-e_\alpha
)^2}(p'_1x'_\alpha -p'_\alpha x'
_1)^2 - {1\over 2(e_1-e_\alpha) }(p'_1x'_\alpha -p'_\alpha x'_1)
(p'_2x'_\alpha -p'_\alpha x'_2) \cr
&+{\sum _\beta}'
{(p'_\alpha x'_\beta -p'_\beta x'_\alpha )^2\over 4(e_\alpha -e_\beta) }]
+
{1\over (u-e_1)}\sum ^{n+1}_{\alpha =3}[{-1\over 4(e_1-e_\alpha
)}_2(p'_1x'_\alpha
-p'_\alpha x'_1)^2\cr
& + {1\over 2(e_1-e_\alpha )}(p'_1x'_\alpha -p'_\alpha x'_1)
(p'_2x'_\alpha -p'_\alpha x'_2)]\cr
&=\sum ^{n+1}_{\alpha =3}{H_\alpha \over u-e_\alpha } + {Z\over (u-e_1)^2} +
{Y\over u-e_1}
\endalign$$
where $Y=\sum ^{n+1}_{\alpha =3}H_\alpha$.

The analysis presented so far could have been deduced from Kuznetsov \cite{4}
where the
double root is essentially contained in the $s$ systems of type $C$ on the real
hyperboloid. Furthermore, the threefold root is contained in Kuznetsov's type D
systems.
The question we now answer is how to use these techniques on the
case of ellipsoidal coordinates corresponding to multiply degenerate roots. For
this we
use the limiting procedures developed by Kalnins, Miller and Reid \cite{5}. We
recall that the
process of using these limiting procedures amounts to altering the elementary
divisors of the two quadratic forms
$$\sum ^{n+1}_{\alpha =1}{x^2_\alpha \over u-e_\alpha} =0,\quad
\sum ^{n+1}_{\alpha =1}x^2_\alpha  = c^2.$$

\proclaim{Theorem 1} Let $u_i$ be the generic ellipsoidal coordinates
coordinates on
the $n$ sphere viz.
$$a_\alpha x^2_\alpha  =
{c^2\Pi ^{n}_{j=1}\ (u_j-e_\alpha )\over \Pi _{\beta \neq \alpha }(e_\beta
-e_\alpha) },
\quad \alpha  = 1,\cdots,n+1 \tag2.18
$$
with corresponding infinitesmal distance
$$ds^2=-\frac14\sum ^{n}_{i=1} \Pi _{j\neq i}(u_i-u_j)
{(du_i)^2\over \Pi ^{n+1}_{j=1}(u_i-e_j)}
$$
and coordinate curves
$$\sum ^{n+1}_{\alpha =1}{a_\alpha x^2_\alpha \over u-e_j} =0,\quad
\sum ^{n+1}_{\alpha =1}a_\alpha x^2_\alpha  = c^2. \tag2.19
$$
Then the  degenerate ellipsoidal coordinates
having the infinitesimal distance
$$ds^2=-\frac14\sum ^{n}_{i=1} [\Pi _{j\neq i}(u_i-u_j)]
{(du_i)^2\over \Pi ^p_{j=1}(u_i-e_j)^{N_j}} \tag2.20
$$
can be obtained from generic ellipsoidal coordinates via the transformations
$$\align
e^J_j&\rightarrow  e^J_1+^J\epsilon ^1_{j-1},\ j=1,\cdots,N_J,\ J=1,\cdots,p
\cr
p^J_j&\rightarrow p^J_1+\sum ^{N_J}_{i=2}{}^J\epsilon ^{i-1}_{j+1-i} p^J_i\cr
x^J_j&\rightarrow x^J_1+\sum ^{N_J}_{i=2}{}^J\epsilon ^{i-1}_{j+1-i} x^J_i
\tag2.21
\endalign$$
where
$${}^J\epsilon ^{i-1}_{j+1-i}=
\Pi ^i_{\ell =2}(^J\epsilon ^1_{j-1}-^J\epsilon ^1_{\ell-2} ),\ \
a^J_j =1/[
\Pi _{k\neq j}(^J\epsilon ^1_{j-1}-^J\epsilon ^1_{k-1} )],\ k=1,\cdots,N_J
$$
and $N_1+\cdots+N_p=n+1$. (We require ${}^J\epsilon_0^i=0$ and take the limit
as the ${}^J\epsilon_h^1\rightarrow 0$ for $h=1,\cdots,N_J-1$.) In particular,
$$
{\bold L}(u) =
\sum ^p_{J=1}\sum ^{N_J}_{j=1} {{\bold z}^J_{j}\over (u-e_J)^{N_J
-j+1}}\tag2.22
$$
where
$$
{\bold z}^J_j=<{1\over 4} \sum _i(p^J_ip^J_{j+1-i}+x^J_ix^J_{j+1-i}),{i\over 4}
\sum _i(p^J_ip^J_{j+1-i}-x^J_ix^J_{j+1-i}),{i\over 2} \sum _ip^J_ix^J_{j+1-i}>.
$$
The $(z^J_j)_\ell $ satisfy the Poisson bracket relations
\medskip
$$\{(z^J_j)_\ell ,(z^{J'}_k)_m\}=-\delta _{JJ'}(z^J_{j+k-N_J})_s
\epsilon _{lms} \tag2.24
$$
and we also have
$$
{\bold L}\cdot{\bold L}(u) = \frac14\sum  (x^J_j p^L_k - x^L_k p^J_j)(x^J_m
p^L_n - x^L_n
p^J_m){1\over (u-e_J)^p(u-e_L)^q} \tag2.25
$$
where
$1\leq p\leq N_J,1\leq q\leq N_L,1\leq j,m\leq N_J,1\leq k,n\leq N_L$, subject
to $j+m=N_J+2-p,k+n=N_L+2-q$ and the summation extends over the indices
$J,L,p,q,j,m,k,n$ subject to these restrictions.
\endproclaim

{}From expressions (2.24) we can verify that relations (2.12) are satisfied, so
that
$$\{{\bold L}\cdot{\bold L}(u),{\bold L}\cdot{\bold L}(v)\}=0.$$
The  constants of the motion can be read off from the
partial fraction decomposition of (2.25). This  clearly
illustrates the compactness of the $r$ matrix formulation for the operators
describing the integrable systems examined so far.
\medskip
In dealing with the case of Euclidean space $E_{n{\bold C}}$ the most
transparent way
to proceed
is as follows. The generic ellipsoidal coordinates in $n$ dimensional complex
Euclidean space are given by \cite{1,9}
$$
x^2_\alpha = {\Pi ^n_{j =1}(u_j-e_\alpha )\over \Pi _{\beta \neq \alpha
}(e_\beta -e_
\alpha )},\quad  j,\alpha ,\beta =1,\cdots,n\tag2.26
$$
with coordinate curves
$$
\sum ^n_{\alpha =1}{x^2_\alpha \over (u-e_\alpha)} =1,\ u=u_j,\
j=1,\cdots,n.\tag2.27
$$

Proceeding in analogy with (2.2)-(2.10) we can obtain the $r$ matrix algebra.
The
corresponding ${\bold L}(u)$ operator is
$$
{\bold L}(u)= \sum ^n_{\alpha =1}{{\bold s}_\alpha \over u-e_\alpha } +
{1\over 4}\left(\matrix -1\\ 1\\ 0\endmatrix\right).\tag2.28
$$
(Indeed the equation $L^1(u)+iL^2(u)=0$ is just (2.27). Moreover it is obvious,
due to the fact
that expressions (2.11) satisfy (2.12), that expressions (2.28) also satisfy
(2.12).) The conjugate
variables $v_j$ are defined by $v_j=-iL^3(u_j)$ and they must satisfy the
canonical relations
(2.4).
The integrals of motion $H_\alpha $ are determined from
$$
{\bold L}^2(u) = \sum ^n_{\alpha =1}{H_\alpha \over u-e_\alpha }\tag2.29
$$
where
$$
H_\alpha  = 2{\sum ^{n}_{\beta =1}}'{({\bold s}_\alpha ,{\bold s}_\beta )
\over e_\alpha -e_\beta } -
{1\over 2}(s^1_\alpha +s^2_\alpha ) ={1\over 4}({\sum ^{n}_{\beta
=1}}'{M^2_{\alpha \beta }\over e_\alpha -e_\beta
} - p^2_\alpha )
$$
with $\sum _\alpha H_\alpha = -{\bold p}^2/4$. The separation equations are of
the form
$$
v^2_j+{\bold L}^2(u_j)=0,\quad j=1,\cdots,n,
$$
as in (2.8).

\subheading{3. Cyclidic Coordinates} Associated with the separation of
variables problem for
the Hamilton Jacobi equation (1.1) with $E\neq 0$ is the
corresponding $E=0$ problem. In this case the equation is
$$
\sum ^n_{\alpha ,\beta  =1} g^{\alpha \beta }p_\alpha p_\beta =0,\quad p =
{\partial W\over \partial x_\alpha },\quad \alpha  =1 ,\cdots,n\tag3.1
$$
and we consider only complex Euclidean space.
While it is true that all the coordinate systems discussed for $E_{n{\bold C}}$
with $E\ne 0$ will provide a separation of variables of this equation, there
are coordinates
that provide an additive separation of variables {\it only} when $E=0$.
This is related to the fact that the $E=0$ equation    admits a conformal
symmetry algebra \cite{1,11,13}.

The  most convenient way to proceed  is to introduce hyperspherical
coordinates\hfill\break
 $\{x_1,
\cdots,x_{n+2}\}$,

$$
x_1=t^2(\sum ^n_{j=1}z^2_j - 1),\ x_{2}=it^2(\sum ^n_{j=1}z^2_j + 1),\
x_{k+2}=2z_kt^2,\ k=1,\cdots,n,
$$
related to the usual Cartesian coordinates $\{z_1,\cdots,z_n\}$
according to
$$
z_k=x_{k+2}/(-x_1-ix_2), \quad k=1,\cdots,n\qquad \sum_{j=1}^{n+2}x_j^2=0.
$$

We consider the system of \S 1 in $n+2$ dimensions, where ${\bold J} = 0$ and
the $c_\alpha
=0$.
The general (separable) cyclidic coordinates are
specified by
$$
\Omega = \sum ^{n+2}_{j=1}{x^2_j\over \lambda -e_i} =0,\ \Phi  =
\sum ^{n+2}_{j=1}x^2_j =0,\ \lambda =u _1,\cdots,u _n,\ \
e_k\neq  e_j.
$$
Furthermore,
$$
\sigma x^2_\alpha=
{\Pi ^n_{j=1}(u _j-e_\alpha )\over
\Pi _{\beta \neq \alpha }(e_\beta -e_\alpha) },\quad
\sigma =-[\sum ^{n+2}_{j=1}e_jx^2_j]^{-1}.
$$
The Hamilton Jacobi equation is given by
$$
{\bold J}^2=-\sigma ^2\sum ^n_{j=1}\left(\frac{\partial W}{\partial
u_j}\right)^2
{\Pi ^{n+2}_{\gamma =1}(u_j-e_\gamma )\over \Pi _{i\neq j}(u_j-u_i)} =0.\tag3.2
$$

The quadratic forms $\Omega $
and $\Phi $ have elementary divisors $[11\cdots 1]$, see \cite{5,11}. It is
known that the geometry of
these fourth order coordinate curves is unchanged under birational
transformations of the form
$$
e_k\rightarrow
{\alpha e_k+\beta \over \gamma e_k+\delta }, \quad
u_j\rightarrow
{\alpha u_j+\beta \over \gamma u_j+\delta }, \quad
\lambda\rightarrow
{\alpha \lambda+\beta \over \gamma \lambda+\delta },
$$
for
$\alpha \delta -\beta \gamma  \neq
0$ and $ k=1,\cdots,n+2,\  j=1,\cdots,n$.

Now  we can mimic the exposition given for the Gaudin magnet
integrable systems using the hyperspherical coordinates and the Poisson bracket
$$
\{F(x_i ,p_j),G(x_i ,p_j)\}_h = \sum ^{n+2}_{k=1}
({\partial F\over \partial p_k} {\partial G\over \partial x_k}-
{\partial F\over \partial x_k}{\partial G\over \partial p_k}),\tag3.3
$$
the $x_i,p_j$ now being regarded as independent. The analysis then proceeds
much as
in the construction (2.2)-(2.12), but with $n$ replaced by $n+1$;
indeed the coordinates are defined as zeros of the off diagonal element
$$
B(u)=\sum ^{n+2}_{\alpha =1}{x^2_\alpha \over u-e_\alpha }
$$
subject to the restriction $\Phi =0$. The  crucial difference is
that ${\bold J} = 0$. The expressions for $p_\alpha $ and $M_{\alpha \beta }$
are altered by a
factor $\sigma $:
$$
p_\alpha =2\sigma x_\alpha[ \sum ^n_{j=1}{v_j\over e_\alpha-u_j}
{\Pi ^n_{\gamma =1}(u_j-e_\gamma )\over \Pi _{i\neq j}(u_j-u_i)}]
$$
which together with $x^2_\alpha $ in terms of $u_j$ gives $C(u)$ in the new
variables. Another useful formula is
$$
M_{\alpha \beta } = 2\sigma x_\alpha x_\beta
(e_\alpha -e_\beta ) \sum ^n_{j=1} v_j
{{\hat\Pi} ^{n+2}_{\gamma =1}(u_j-e_\gamma )\over \Pi _{i\neq j}(u_j-u_i)} .
$$

In fact the Poisson bracket $\{ ,\}_h$ can be identified with the
Poisson bracket $\{$ , $\}$ for functions defined in the $n$ dimensional space
spanned by $z_1,\cdots,z_n$. This can  readily be seen by noting that
$F(z_1,\cdots,z_n,p_{z_1},\dots,p_{z_n})=F(-x_3/(x_1+ix_2),\cdots,-x_{n+2}/(x_1+ix_2),
-(x_1+
ix_2)p_{x_3}+x_3(p_{x_1}+ip_{x_2}),\cdots,-(x_1+ix_2)p_{x_{n+2}}+x_{n+2}(p_{x_1}+ip_{x_2}))$
from
which the equality of the Poisson brackets follows identically. The
infinitesmal
distance for  general cyclidic coordinates is
$$
ds^2=\frac14(x_1+ix_2)^{-2}\left(\sum ^n_{i=1} \Pi _{j\neq i}(u_i-u_j)
{(du_i)^2\over \Pi ^{n+2}_{j=1}(u_i-e_j)}\right). \tag3.4
$$
We denote the coordinates defined by this graph as
$$[E_{n{\bold C}},E=0|e_1|....|e_{n+2}|.
$$
Other coordinates of this type are those corresponding to the graphs
$$(S_{p{\bold C}}|e_1|....|e_{p+1}|\oplus (S_{q{\bold C}}|f_1|....|f_{q+1}|
$$
where $p+q=n+2$. These coordinates are given by
$$\align
\sigma x^2_k&=
{\Pi ^p_{j=1}(u_j-e_\alpha )\over \Pi _{\beta \neq \alpha }(e_\beta -e_\alpha )
},\quad \alpha ,\beta  =1,\cdots,p+1,\ k=1,\cdots,p\tag3.5\cr
\sigma x^2_k&=
{\Pi ^q_{j=1}(v_j-f_\alpha )\over \Pi _{\beta \neq \alpha }(f_\beta -f_\alpha )
},\quad \alpha ,\beta  =1,\cdots,q+1,\ k=p+1,\cdots,p+q+2\cr
\sigma & = -[\sum ^{p+1}_{i=1}e_ix^2_i+\sum ^{q+1}_{j=1}f_jx^2_j]^{-1}.
\endalign$$
We note here that  the $e_i$  the $f_j$ are pairwise distinct, for if they were
not then for any of these quantities which occured with multiplicity more than
1 a birational transformation could transform it to $\infty $ and hence to a
graph corresponding to $E_{n{\bold C}}$.

It can happen just as in the case of generic ellipsoidal integrable systems on
the sphere that some of the $e_i$ in (3.5) are equal to some of the $f_j$ also.
In  this case the rules for obtaining  the corresponding $L$ matrix are
summarised
in the following theorem.

\proclaim{Theorem 2} Denote the generic ellipsoidal coordinates by the graph

$$[E_{n{\bold C}} |e_1|....|e_n],\quad e_i\neq e_j$$
and generic cyclidic coordinates by the graph
$$\{CE_{n{\bold C}}|e_1|.....|e_{n+2}\},\quad e_i\neq e_j.$$
Separable coordinates for the Hamilton Jacobi equation (3.1)
corresponding to  generic graphs with multiplicities
$$[E_{n{\bold C}} |e^{n_1}_1|\cdots|e^{n_p}_p],\quad n_1+\cdots+n_p=n,
$$
$$\{CE_{n{\bold C}}|e^{n_1}_1|\cdots|e^{n_q}_q\},\quad n_1+\cdots+n_q=n+2,$$
\cite{5} can be obtained via the transformations of Theorem 1 applied to the
quadratic
forms
$$\Omega  = \sum ^{m}_{i=1}{a_ix^2_i\over \lambda -e_i} =0,\quad \Phi  =
\sum ^{m}_{i=1}a_ix^2_i =0
$$
where $m=n$ for generic ellipsoidal coordinates in $E_{n{\bold C}}$ and $m=n+2$
for the corresponding cyclidic coordinates.
\endproclaim

For coordinates corresponding to the direct sum of two spheres viz. (3.5)
we can merely apply the result of the previous theorem for spherical
coordinates to each of the pairs of quadratic forms
$$\Omega '_1=\sum ^{p+1}_{\alpha =1}{a_\alpha x^2_\alpha \over u-e_\alpha },
\quad
\Phi '_1= \sum ^{p+1}_{\alpha =1}a_\alpha x^2_\alpha $$
$$\Omega '_2=\sum ^{n+2}_{\alpha =p+2}{a_\alpha x^2_\alpha \over u-e_\alpha
},\quad
\Phi '_2= \sum ^{n+2}_{\alpha =p+2}a_\alpha x^2_\alpha. $$

Indeed the freedom to subject the coordinates $u_j$ and the $e_i$ to birational
transformations in the expressions for generic cyclidic coordinates allows us
to let $e_1\rightarrow  \infty$.
(In this particular case  the resulting coordinates can be identified with
generic elliptical coordinates on the $n$ sphere.)  The process
described in Theorem 1 enables one to pass from the elementary divisors
[11....1] to
$[N_1,N_2,....,N_p] ,N_i\geq 1 ,i=1,...,p$, see \cite{5,11}. It is then
possible (via Theorem 2)
to take
$e_1\rightarrow  \infty $ in which case $[{\overset \infty \to
N}_1,N_2,....,N_p]$
corresponds to the various generic coordinate systems in
Euclidean space if $N_1>1$. To illustrate how this works for the Gaudin
XXX magnet model
consider the quadratic forms $\hat \Omega  ,\hat \Phi $
corresponding to elementary divisors $[21\cdots1]$ viz.
$$
\hat\Omega ={x^2_1\over (\lambda -e_1)^2} + {2x_1x_2\over \lambda -e_1} +
\cdots
+ {x^2_{n+2}\over \lambda -e_{n+2}}=0,\
\hat \Phi =2x_1x_2+ x^2_3+\cdots+ x^2_{n+2}=0.
$$
Putting  $\lambda \rightarrow 1/\lambda,e_i\rightarrow 1/e_i$ we find (with the
use
of $\hat \Phi=0$)
$$
\hat\Omega =\lambda ^2[{e^2_1x^2_1\over (e_1-\lambda )^2} + \sum ^{n+2}_{k=3}
{(e_1-e_k)\over (e_1-\lambda )} {x^2_k\over (e_k-\lambda )} ],
$$
or, in the limit as $e_1\rightarrow\infty$,
$$
x_1^2-\sum_{k=3}^{n+2}\frac{x_k^2}{\lambda-e_k}=0.
$$
Now if we perform the limiting procedure (2.13) - (2.15) on ${\bold L}(u)$ (in
$n+2$
dimensions), let $u\rightarrow 1/u$,  $e_i\rightarrow 1/e_i$,  let
$e_1\rightarrow\infty$ and then evaluate at  $p_1= 0, x_1=  1$ we find (up to
a common factor $u$)
$$
{\bold L}(u)= \sum ^{n+2}_{\alpha =3}{{\bold s}_\alpha \over u-e_\alpha } +
{1\over 4}\left(\matrix -1\\1\\0\endmatrix\right).
$$
Here, we have made use of the fact that ${\bold J}=0$.
This agrees with (2.28) and with the ${\bold L}(u)$ operator given by Kuznetsov
\cite{4},
and corresponds to elementary divisors $[{\overset \infty \to 2},1,\cdots, 1]$.

This process can
 be generalised for quadratic forms corresponding to elementary divisors
$[{\overset \infty \to N},1\cdots,1]$, $N>1$.  We have that
$$\hat \Omega =u ^{N-2}x^2_1 + u ^{N-3}2x_1x_2       +\sum _ix_ix_{N-i}
- \sum ^{n+2}_{k=N+1}{x^2_k\over {u -e_k}},\
\hat\Phi =\sum _{i=1}^N x_ix_{N+1-i}+\sum_{k=N+1}^{n+2}x_k^2.
$$
The corresponding ${\bold L}(u)$ operator is
$$
{\bold L}(u)=\sum ^{n+2}_{\alpha=N+1} {{\bold s}_\alpha\over u-e_k}  + {\bold
s}_1
$$
where
$$\align
s^1_1&= {1\over 4}[\hat\Omega (0,p_2,...,p_{n+2})-\hat\Omega
(1,x_2,...,x_{n+2})],\cr
s^2_1&= {1\over 4}[\hat\Omega (0,p_2,...,p_{n+2})+\hat\Omega
(1,x_2,...,x_{n+2})],\cr
s^3_1&= {1\over 2}[S'(0,p_2,....,p_{n+2};1,x_2,....,x_{n+2})],\quad
S'=\sum ^N_{j=2}u ^{N-j}\sum _{\ell +m=j}x_\ell p_m.
\endalign$$

\subheading{4. Branching Rules for the Construction of Orthogonal Non
Generic Complex Integrable Systems on  $S_{n{\bold C}}$ and $E_{n{\bold C}}$}
To deal with the non generic separable coordinate systems in $E_{n{\bold C}}$
and
$S_{n{\bold C}}$ we must combine coordinate systems for both manifolds
\cite{5}.
(See \cite{13,14} for tabulations of all cases for small values of $n$.) The
branching laws for graphs on these manifolds are summarised below
$$
  \matrix (S_{n{\bold C}}|&\cdots&|e_i|&\cdots|\\
                                    &       &\downarrow&\\
                                    &       &S_{p_i{\bold C}}&\endmatrix \tag1
$$
$$
  \matrix (S_{n{\bold C}}|&\cdots&|e^{\lambda_i}_i|&\cdots|\\
                                    &       &\downarrow&\\
                                    &       &E_{p_i{\bold C}}&\endmatrix,\ \
\lambda_i>1 \tag2
$$
$$
  \matrix (E_{n{\bold C}}|&\cdots&|e_i|&\cdots|\\
                                    &       &\downarrow&\\
                                    &       &S_{p_i{\bold C}}&\endmatrix \tag3
$$
$$
 \matrix (E_{n{\bold C}}|&\cdots&|e^{\lambda_i}_i|&\cdots|\\
                                    &       &\downarrow&\\
                                    &       &E_{p_i{\bold C}}&\endmatrix,\ \
\lambda_i>1.\tag4
$$

As an example consider the coordinate system given by the graph on
$S_{4{\bold C}}$,
$$
\matrix (S_{2{\bold C}}&|e^2_1|&e_5|\\
                        &\downarrow&\\
        [E_{2{\bold C}}&|f_3|  &f_4|&\endmatrix.
$$

The coordinates for this graph can be obtained from those of the generic graph
$$
(S_{4{\bold C}}|e^2_1|e_3|e_4|e_5|$$
via the limiting transformations
$$\align
e_j&=e_1+\epsilon +\epsilon ^2f_j,\quad j=3,4,\quad f_3\neq f_4 \cr
u_j&=e_1+\epsilon +\epsilon ^2\bar u_j,\quad j=3,4
\endalign$$
where $\epsilon \rightarrow 0$.

The corresponding coordinates are then  given implicitly by
$$\align
-x^2_1 &= {(u_1-e_1)(u_2-e_1)\over e_5-e_1},\ \ x^2_5 =
{(u_1-e_5)(u_2-e_5)\over (e_1-e_5)^2}\cr
-2x_1x_2 &= - {(u_1-e_1)\over e_5-e_1} - {(u_2-e_1)\over e_5-e_1} +
{(u_1-e_1)(u_2-e_1)\over (e_5-e_1)^2}\cr
& - {(u_1-e_1)(u_2-e_1)\over e_5-e_1}
(f_3+f_4-\bar u_3-\bar u_4)\cr
-x^2_3 &={(u_1-e_1)(u_2-e_1)\over (e_5-e_1)}
{(\bar u_3-f_3)(\bar u_4-f_3)\over (f_3-f_4)},\cr
-x^2_4 &={(u_1-e_1)(u_2-e_1)\over (e_5-e_1)}
{(\bar u_3-f_4)(\bar u_4-f_4)\over (f_4-f_3)}.
\endalign$$

Using the nomenclature given previously we have that ${\bold L}(u)$ has the two
different forms
$$\align
{\bold L}_1(u)&={{\bold z}_1\over (u-e_1)^2}+ {{\bold z}_2\over u-e_1} +
 {{\bold z}_3 + {\bold z}_4\over u-e_1}+ {{\bold z}_5\over u-e_5},\ \ u =
u_1,u_2\cr
{\bold L}_2(u)&= {\bold z}_1+ {{\bold z}_3\over u-f_3} + {{\bold z}_4\over
u-f_4},\
\ u = \bar u_3,\bar u_4.
\endalign$$
As usual, the separation variables are the zeros of the equation
$L^1_\lambda(u)
+iL^2_\lambda(u)=0$ with $u=u_1,u_2$ when $\lambda=1$ and $u={\bar u_3}, {\bar
u_4}$ when
$\lambda=2$. Each of the ${\bold L}_\lambda$ separately satisfy (2.12) for
$\lambda
=1,2$. In addition we have
$$\{{\bold L_1}\cdot{\bold L_1}(u),{\bold L_2}\cdot{\bold L_2}(v)\}=0.$$

This example illustrates how to derive  the substitutions that enable the
various
branching laws to be obtained from a generic form. For the sphere
$S_{n{\bold C}}$ and generic coordinates
$$
(S_{n{\bold C}}|...|e^{\lambda +q-s}_0|e_1|\cdots|e_s|\cdots|\qquad q>s,\
\lambda >1,
$$
if we make the substitutions
$$\align
e_k&=e_0+\epsilon ^{q-s+1}+\epsilon ^{\lambda +q-s}f_k,\ \ k=1\cdots s,\cr
u_j&=e_0+\epsilon ^{q-s+1}+\epsilon ^{\lambda +q-s}\bar u_j,\ \ j=1\cdots q
\endalign$$
where $\epsilon \rightarrow 0$,
then the graph illustrated transforms into
$$
\matrix (S_{n{\bold C}}|&\cdots&\cdots&|e^\lambda_0|&\cdots|\\
                        &      &      &\downarrow   &       \\
        [E_{q{\bold C}}|&f_1|  &\cdots&\cdots       &|f_s| \endmatrix.
$$

The remaining branching rule is obtained from a graph of the type
$$(S_{n{\bold C}}|...|e_0|e_1|....|e_s|.
$$
By means of the substitutions
$$
e_j=e_0+\epsilon (f_0-f_j),\ j=1,\cdots,p+1\qquad
u_k=e_0+\epsilon (\bar u_k-f_0),\ k=1,\cdots,p
$$
we  obtain the coordinate system coming from the graph
$$
\matrix (S_{n{\bold C}}|&\cdots&\cdots&|e_0|&\cdots|\\
                        &      &      &\downarrow   &       \\
        (S_{p{\bold C}}|&f_0|  &\cdots&\cdots       &|f_p| \endmatrix.
$$

The corresponding substitutions  for the analogous Euclidean space
branching rules are essentially identical. To completely specify the coordinate
systems on these manifolds we need a few more substitution rules. Firstly
consider the graph
$$
[E_{p{\bold C}}|h_1|....|h_p|\  +\  [E_{q{\bold C}}|f_1|....|f_q|.
$$
This graph can be obtained from the generic graph for $E_{(p+q){\bold C}}$
via the substitutions
$$\align
u_i&= {K_1\over \epsilon } + \bar u_i,\ e_i={K_1\over \epsilon } + h_i,
\ i=1,\cdots,p\cr
u_i&= {K_2\over \epsilon } + \bar u_i,\ e_i={K_2\over \epsilon } + f_{i-p},\
i=p+1,\cdots,p+q,\quad K_1\ne K_2.
\endalign$$
The graph $[E_{q{\bold C}}|h_1|....|h_q|$, $q<n$ can be obtained from
$[E_{n{\bold C}}|h_1|....|h_n|$
 via the substitutions
$$\align
u_i&=e_1+\epsilon ^{q-n-1}+\epsilon ^{q-n}\bar u_i,\ i=1,\cdots,n \cr
e_j&=e_1+\epsilon ^{q-n-1}+\epsilon ^{q-n}h_j,\ j=1,\cdots,q\quad
e_k=h_k,\ k=q+1,\cdots,n.
\endalign$$

Given these substitution rules all the corresponding graphs for $E_{n{\bold
C}}$
and $S_{n{\bold C}}$ can be constructed together with their corresponding $L$
matrices.

\subheading{5.  Quantum Integrable Systems on Complex Constant Curvature
Spaces and the
Quantum Gaudin Magnet}
To deal with the quantum version of this description of separation of variables
we consider the Schr\"odinger or Helmholtz equation [3].
$${\Cal H}\Psi ={1\over \sqrt{ g}}
{\partial \over \partial y_\alpha }[\sqrt{ g}\ g^{\alpha \beta }
{\partial \over \partial y_\beta }]\Psi =E\Psi, \tag5.1
$$
\medskip
\noindent
where $E\neq 0$ for the moment. Separation of variables means (roughly) the
solution of
this equation of the form
$$\Psi =\Pi ^n_{\alpha =1}\psi _\alpha (y_\alpha ;h_1,\cdots,h_n)\tag5.2$$
where the quantum numbers $h_j $ are the eigenvalues of mutually commuting
operators
$$\align
{\Cal A}_i=&\sum ^n_{\alpha ,\beta =1}a^{(j)}_{\alpha \beta }{\partial ^2\over
\partial y_\alpha \partial y_\beta }+ \sum ^n_{\beta =1}b_\beta
{\partial \over \partial y_\beta },\quad j=1,\cdots,n \tag5.3  \\
[{\Cal A}_i,{\Cal A}_j]=&0,\quad i,j=1,\cdots,n,\ {\Cal A}_n={\Cal H},
\ {\Cal A}_j\Psi=h_j\Psi.
\endalign$$

Furthermore these operators can be represented as symmetric quadratic elements
in the enveloping algebra of the symmetry algebra of the equation (5.1) in the
case of $S_{n{\bold C}}$ and $E_{n{\bold C}}$. The standardised representations
of
these symmetry algebras are
$$\align SO(n+1):&\ {\Cal M}_{\alpha \beta }=\hat x_\alpha \hat p_\beta -\hat
x_\beta
\hat p_\alpha,\ \hat p_\alpha ={\partial \over \partial x_\alpha }
,\ \alpha ,\beta  =1,\cdots,n+1.\\
&[{\Cal M}_{\alpha \beta } ,{\Cal M}_{\gamma \delta }]
=\delta _{\alpha \gamma }{\Cal M}_{\delta \beta }+\delta _{\alpha \delta }%
{\Cal M}_{\beta \gamma }+\delta _{\beta \gamma }{\Cal M}_{\alpha \delta }+%
\delta _{\beta \delta }{\Cal M}_{\gamma \alpha }\tag5.4  \\
E(n):&\ {\Cal M}_{\alpha \beta }
,\ {\Cal P}_\gamma ={\partial \over \partial x_\gamma }\\
&[{\Cal M}_{\alpha \beta } ,{\Cal P}_\gamma ]
=\delta _{\beta \gamma }{\Cal P}_\alpha -\delta _{\alpha \gamma }{\Cal P}_
\beta ,\ [{\Cal P}_\alpha  ,{\Cal P}_\beta ] =0\tag5.5
\endalign$$
where $[ , ]$ is the commutator bracket. Much of the analysis goes through as
it
did in the classical case with, of course, some critical differences. For the
quantum  Gaudin magnet one considers the sum of rank 1 Lie algebras
$ {\Cal A} = \oplus ^m_{\alpha =1} so_\alpha (3)$ where the generators of the
algebra
satisfy the commutation relations (1.6)  and the inner product is defined as in
\S1. The Casimir elements of ${\Cal A}$ have the form
$$({\bold s}_\alpha  ,{\bold s}_\alpha  )=k_\alpha (k_\alpha +1)\tag5.6$$
where $k_\alpha$ is a constant when the generators of $so_\alpha (3)$ determine
an irreducible representation,\cite{12}. The quantum Gaudin magnet is  the
quantum integrable
Hamiltonian system on ${\Cal A}$ given by $m$ commuting integrals of motion :
$$
{\Cal H}_\alpha =2\ {\sum ^{m}_{\beta =1}}'{({\bold s}_\alpha
,{\bold s}_\beta )\over e_\alpha  -e_\beta },\quad [{\Cal H}_\alpha
,{\Cal H}_\beta ]=0.
$$
This is the $m$ {\it site $SO(3,{\bold C})$-XXX quantum Gaudin
magnet}, \cite{3}. These integrals (operators) satisfy
$$
\sum ^m_{\alpha =1}{\Cal H}_\alpha =0
,\quad \sum ^m_{\alpha =1}e_\alpha {\Cal H}_\alpha  = {\bold  J}^2 -
\sum ^m_{\alpha =1}k_\alpha (k_\alpha +1)\tag5.7
$$
where ${\bold  J} = \sum ^m_{\alpha =1}{\bold s}_\alpha $ is the total momentum
operator. The complete set of commuting operators consists of ${\Cal H}_\alpha
,{\bold  J}^2$ and ${\Cal J}^3$. The integrals are generated by the $2\times 2$
operator $L(u)$ given by (1.11)   understood in the operator sense. The quantum
determinant is
$$ \text{q-det} L(u)=-A(u)^2- {1\over 2}\{B(u),C(u)\}
= -\sum ^m_{\alpha =1}{{\Cal H}_\alpha \over u - e_\alpha } -
\sum ^m_{\alpha =1}{k_\alpha (k_\alpha +1)\over (u - e_\alpha)^2 }\tag5.8
$$
with $L$-operator satisfying the $r$ matrix algebra
$$
[ {\overset 1 \to L}(u), {\overset 2 \to L}(v)]=\frac{i}{u-v}
[P, {\overset 1 \to L}(u)+ {\overset 2 \to L}(v)],\quad i=\sqrt{-1},\tag5.9
$$
(compare with (1.13)).
The total  momentum ${\bold  J}$ has components
$${\Cal J}^1= {1\over 4}(\hat {\bold p}\cdot \hat {\bold p} +
\hat {\bold x}\cdot \hat {\bold x}),\
{\Cal J}^2= {i\over 4}(\hat {\bold p}\cdot \hat {\bold p} -
\hat {\bold x}\cdot \hat {\bold x}),\ {\Cal J}^3=
{i\over 4} (\hat {\bold p}\cdot \hat {\bold x}+\hat {\bold x}\cdot \hat {\bold
p})\tag5.10
$$

The Lie symmetries ${\Cal M}_{\alpha \beta }$ and ${\bold  J}$ form the direct
sum $so(m)\oplus so(3)$ with commutation relations
$$[{\Cal M}_{\alpha \beta } ,{\Cal J}^i]=0
,\quad [{\Cal J}^i,{\Cal J}^j]=i\epsilon _{jk\ell }{\Cal J}^\ell,\tag5.11
$$
subject to the constraints
$${\bold  J}^2= {1\over 4}\sum _{\alpha <\beta } {\Cal M}^2_{\alpha \beta }
+ {1\over 16} m(m-4),\quad
2({\Cal J}^1+i{\Cal J}^2)= \hat {\bold x}\cdot \hat{\bold x}.\tag5.12
$$

Considering again separation of variables using the coordinates of the
irreducible block (2.1), the separation variables are defined, as before, as
the zeros
of the off diagonal elements $B(u)$ of the $L$ matrix. The $q$-determinant is
the generating function of the commuting integrals of motion
$$[\text{q-det} L(u), \text{q-det} L(v)]=0.\tag5.13
$$
On the $n$ sphere the algebra ${\Cal A}$ is realised by taking the canonical
operators
$$s^1_\alpha = {1\over 4} (\hat {  p}^2_\alpha +\hat {  x}^2_\alpha ),\quad
s^2_\alpha = {i\over 4} (\hat {  p}^2_\alpha -\hat {  x}^2_\alpha ),\quad
s^3_\alpha =
{i\over 4} \{\hat p_\alpha ,\hat x_\alpha \},\ \alpha =1,\cdots,n+1\tag5.14
$$
where $({\bold s}_\alpha , {\bold s}_\alpha)  = -{3\over 16}$. Furthermore
$$({\bold s}_\alpha  , {\bold s}_\beta  )= {1\over 8}({\Cal M}^2_{\alpha \beta
} +
{1\over 2})\tag5.15
$$
and this establishes the relationship between the ${\bold s}_\alpha $ of ${\Cal
A}$ and
the ${\Cal M}_{\alpha \beta }$ of $so(n+1)$. The integrals    transform into
the
family of integrals
$${\Cal H} = \sum ^{n+1}_{\alpha =1}h_\alpha {\Cal H}_\alpha  = {1\over 4}
\sum _{\alpha <\beta }
{h_\alpha -h_\beta \over e_\alpha -e_\beta }({\Cal M}^2_{\alpha \beta } +
{1\over 2})\tag5.16
$$
and the coordinates are given by
$$  B(u_j)=0:\qquad \sum ^{n+1}_{\alpha =1}{x^2_\alpha \over u - e_\alpha } =
0.
$$
For each $u_j$ variable there is defined the conjugate variable (operator)
$$v_j=-i A(u_j)= {1\over 4}
\sum ^{n+1}_{\alpha =1}{1\over u_j-e_\alpha } \{\hat p_\alpha ,\hat x_\alpha
\}.\tag5.17
$$
Separation of variables is then the process of changing to the new variables
$v_j,u_i,c$ and ${\Cal J}^3$. In fact we have that
$${1\over 4}\{\hat p_\alpha ,\hat x_\alpha \}=
{\hat x^2_\alpha \over c}[-i{\Cal J}^3-\sum ^{n}_{j=1}
{1\over e_\alpha -u_j}D_jv_j],\quad D_j=-\frac{\Pi_{\alpha
=1}^{n+1}(u_j-e_\alpha)}{\Pi_{i\ne j}(u_j-u_i)}.\tag5.18$$
The separation equations can be obtained by substituting $u=u_j$ into the
q-det$L(u)$,  making the choice
$$v_j=i( {\partial \over \partial u_j}+ \frac14
\sum ^{n+1}_{\alpha =1}{1\over u_j-e_\alpha }),\tag5.19
$$
and looking for the solutions of the spectral problem
${\Cal H}_\alpha \Psi =h_\alpha \Psi$. The separation equations are then
$${d^2\over du^2_j}\psi _j +
{1\over 2}(\sum ^n_{\alpha =1}{1\over u_j-e_\alpha }){d\over du_j}\psi _j =
\sum ^{n+1}_{\alpha =1}{h'_\alpha \over {u_j-e_\alpha} }\psi _j\tag5.20
$$
where
$$\Psi  =\Pi ^{n+1}_{j=1}\psi _j(u_j),\quad h'_\alpha  = h_\alpha  + {1\over 8}
\sum _{\beta \neq \alpha } {1\over {e_\alpha -e_\beta} }.
$$

These are the separation equations for the Helmholtz equation in generic
ellipsoidal coordinates. Note that for this choice of $v_j$ we have taken
$({\bold s}_\alpha , {\bold s}_\alpha) =-3/16$. The $L$ operator is given in
direct analogy
with the classical case,
$${\bold  L}(u)= \sum ^{n+1}_{\alpha =1}{{\bold s}_\alpha \over u-e_\alpha
}.\tag5.21$$

The treatment of generic ellipsoidal coordinates on $E_{n{\bold C}}$ follows
similar lines, with the $L$ operator given in this case by

$${\bold  L}(u)= \sum ^n_{\alpha =1}{{\bold s}_\alpha \over u-e_\alpha } +
{1\over 4}\left(\matrix -1\\1\\0\endmatrix\right).\tag5.22
$$
The limiting procedures described in the classical case work also in the
quantum case. All
coordinate systems that were obtained in the classical case appear again.
 In the case of cyclidic
coordinates we can  adopt the same strategy as in \S3:  we impose the
conditions
${\Cal J} =0$ and proceed as before. The natural setting in this case is again
to
use hyperspherical coordinates. The total  momentum has components as
in  (5.10). If $\hat p_\alpha ,\hat x_\alpha $ are now vectors in
hyperspherical
coordinates then we can derive the standard quantum $r$ matrix algebra as above
with the same formulas   valid. The constraints are now
$2({\Cal J}^1+i{\Cal J}^2)= \hat{\bold  x}\cdot \hat{\bold  x} =0 ,\ { \bold
J}^2= {1\over 8}
n(n-4)$. Coordinates $u_j$ and their conjugate operators $v_j$ can be chosen as
before. In
particular if we make the choice
$$
v_j= {\partial \over \partial u_j}+ \frac14
\sum ^{n+2}_{\alpha =1}\frac{1}{u_j-e_\alpha} \tag5.23
$$
and look for solutions of $H\Psi =0$ of the form $\Psi =\sigma ^{(n-2)/2}
\Pi ^n_{i=1}\psi _i(u_i)$, then we obtain the separation equations    which
coincide with the the equations given by Bocher,\cite{11}, viz.
$$\align
{d^2\over du^2_j}\psi _j + &
{1\over 2}\sum ^{n+2}_{\alpha =1}{1\over u_j-e_\alpha }{d\over du_j}\psi _j =
\\
&{{-1\over 16}(n^2-4)u^n_j-{1\over 16}(2n-n^2)(\sum ^{n+2}_{\alpha =1}e_%
\alpha )u^{n-1}_j+
\sum ^{n-2}_{\beta =0}\lambda _\beta u^\beta _j\over \Pi ^{n+2}_{\gamma
=1}(u_j%
-e_\gamma )}\psi _j
\endalign$$
for suitable $\gamma _\beta$.  Note that the solutions $\Psi$ are not strictly
separable
in this case but are what is termed $R$ separable, i.e., separable to within a
non separable
factor $\sigma^{(n-2)/2}$, \cite{1}.

We conclude this work by pointing out that one can obtain a complete set of
constants of
the motion associated with an orthogonal separable coordinate system
$\{u_1,\cdots u_n\}$
for the
Schr\"odinger equation (5.1) directly from (5.1) itself, \cite{15, 16}. It is
well known that all orthogonal
separable coordinate systems for the Schr\"odinger equation on a Riemannian
space are
obtainable via the St\"ackel construction, e.g., \cite{17, 18}. Thus if
$\{\bold u\}$ is a separable
orthogonal coordinate system for (5.1) there exists an $n\times n$ nonsingular
matrix
$S=\left(S_{\gamma,\beta}(u_\alpha)\right)$ such that
$\partial_{u_\gamma}S_{\alpha\beta}=0$ if $\gamma\ne\alpha$,
and such that the nonzero components of the contravariant metric tensor in the
coordinates
$\{\bold u\}$ are $g^{\alpha\alpha}(\bold u)=T^{n\alpha}(\bold u),\
\alpha=1,\cdots, n$, where $T$ is
the inverse matrix to $S$,
$$
\sum_{\beta=1}^n T^{\alpha\beta}(\bold
u)S_{\beta\gamma}(u_\beta)=\delta^\alpha_\gamma.\tag5.24
$$
The constants of the motion are
$$
{\Cal A}_\beta=\sum_{\alpha=1}^n T^{\beta\alpha}(\bold
u)\left(\partial_{u_\alpha}^2+f_\alpha(u_\alpha)\partial_{u_\alpha}\right),\quad {\Cal A}_n={\Cal
H}.\tag5.25
$$
Here, $ f_\alpha(u_\alpha) = \partial_{u_\alpha} \ln(\sqrt{g}
g^{\alpha\alpha})$. (The fact that
$$
\partial_{u_\alpha u_\beta}\ln(\sqrt{g} g^{\alpha\alpha})=0\ \text{for} \
\alpha\ne\beta
$$
follows for any space of constant curvature by noting that it is equivalent to
the statement
that the off diagonal elements of the Ricci tensor must vanish for an
orthogonal coordinate
system: $R_{\alpha\beta}=0$, $\alpha\ne\beta$, \cite{17}.) One can show that
$[{\Cal A}_\alpha,{\Cal
A}_\beta]=0$, \cite{17,  18}. Furthermore, if $
\Psi =\Pi ^n_{\alpha =1}\psi _\alpha (y_\alpha;h_1,\cdots,h_n)$ satisfies the
{\it
separation equations}
$$
\partial_{u_\alpha}^2\Psi+f_\alpha(u_\alpha)\partial_{u_\alpha}\Psi=\sum_{\beta=1}^n
S_{\alpha\beta}(u_\alpha)h_\beta\Psi,\ \alpha=1,\cdots,n\tag5.26
$$
then
$$
{\Cal A}_\beta\Psi=h_\beta\Psi,\quad \beta=1,\cdots,n.\tag5.27
$$

Now fix $\gamma$, $1\le \gamma\le n$, and denote by $\bold y$ the coordinate
choice $${\bold
y}=(u_1,\cdots,u_{\gamma-1},\tau, u_{\gamma+1},\cdots,u_n)$$ where $\tau$ is a
parameter.
We see from (5.26) that if $\Psi$ satisfies the separation equations then
$$
\psi_\gamma^{-1}\left(\partial_{u_\gamma}^2\psi_\gamma+f_\alpha(u_\gamma)\partial_{u_\gamma}\psi_\gamma\right)
|_{u_\gamma=\tau}\Psi({\bold u})=\sum_{\beta=1}^n S_{\gamma\beta}(\tau){\Cal
A}_\beta\Psi({\bold u}).\tag5.28
$$
On the other hand, from (5.1) we have
$$
\psi_\gamma^{-1}\left(\partial_{u_\gamma}^2\psi_\gamma+f_\alpha(u_\gamma)\partial_{u_\gamma}\psi_\gamma\right)
|_{u_\gamma=\tau}\Psi({\bold u})=\frac{1}{g^{\gamma\gamma}(\bold y)}\left({\Cal
H}(\bold u)-
{\Cal H}^\gamma(\bold y)\right)\Psi({\bold u})
$$
where
$$
{\Cal H}^\gamma(\bold y)=\frac{1}{\sqrt{g(\bold
y)}}\sum_{\alpha\ne\gamma}\partial_{u_\alpha}\left(
\sqrt{g(\bold y)} g^{\alpha\alpha}(\bold y)\partial_{u_\alpha}\right).
$$
This suggests the operator identity
$$
\frac{1}{g^{\gamma\gamma}(\bold y)}\left({\Cal H}(\bold u)-
{\Cal H}^\gamma(\bold y)\right)=\sum_{\beta=1}^n S_{\gamma\beta}(\tau){\Cal
A}_\beta,\tag5.29
$$
i.e., that the left-hand side of (5.29) is a one-parameter family of constants
of the
motion and that as $\tau$ runs over a range of values and $\gamma=1,\cdots,n$
the full space of
constants of the motion associated with this separable system is spanned,
\cite{15, 16}.

Now (5.29) is equivalent to the conditions
$$
\frac{T^{n\alpha}(\bold u)}{T^{n\gamma}(\bold y)}-\frac{T^{n\alpha}(\bold
y)}{T^{n\gamma}(\bold y)}
(1-\delta^\alpha_\gamma)=\sum_{\beta=1}^n
S_{\gamma\beta}(\tau)T^{\beta\alpha}(\bold u),
$$
and these conditions are easily seen to follow from (5.24) and from
$$
\sum_{\beta=1,\beta\ne\gamma}^n T^{n\beta}(\bold
y)S_{\beta\xi}(u_\beta)+T^{n\gamma}(\bold
y)S_{\gamma\xi}(\tau)=\delta^n_\xi.$$

Similarly, for the classical case an expression for the Hamiltonian analogous
to
(5.29) generates the constants of the motion, \cite{15, 20}.

\enddocument